\documentclass[preprint,letterpaper]{aastex}

\usepackage[pass]{geometry}
\usepackage{graphicx}
\usepackage{afterpage}
\usepackage{color}

\shorttitle{ASYMMETRIC CHROMOSPHERIC RECONNECTION}
\shortauthors{MURPHY \& LUKIN}

\newcommand\dif{\ensuremath{\mathrm{d}}}

\newcommand{\J}{\ensuremath{\mathbf{J}}}
\newcommand{\E}{\ensuremath{\mathbf{E}}}
\newcommand{\B}{\ensuremath{\mathbf{B}}}
\newcommand{\Rini}{\ensuremath{\mathbf{R}^{in}_i}}
\newcommand{\Rcxin}{\ensuremath{\mathbf{R}^{cx}_{in}}}
\newcommand{\Rcxni}{\ensuremath{\mathbf{R}^{cx}_{ni}}}
\newcommand{\vi}{\ensuremath{\mathbf{V}_i}}

\newcommand{\vn}{\ensuremath{\mathbf{V}_n}}
\newcommand{\ptensi}{\ensuremath{\mathsf{P}_i}}
\newcommand{\ptense}{\ensuremath{\mathsf{P}_e}}
\newcommand{\ptensn}{\ensuremath{\mathsf{P}_n}}

\newcommand{\tasym}{\ensuremath{\mathcal{T}}}
\newcommand{\basym}{\ensuremath{\mathcal{B}}}
\newcommand{\nasym}{\ensuremath{\mathcal{N}}}
\newcommand{\fasym}{\ensuremath{\mathcal{F}}}
\newcommand{\ptot}{\ensuremath{P_\mathrm{tot}}}

\newcommand{\yhat}{\ensuremath{\hat{\mathbf{y}}}}
\newcommand{\zhat}{\ensuremath{\hat{\mathbf{z}}}}
\newcommand{\thick}{\ensuremath{\lambda_\psi}}

\defcitealias{leake:partial1}{L12}
\defcitealias{leake:partial2}{L13}

\begin{document}

\title{ASYMMETRIC MAGNETIC RECONNECTION IN WEAKLY IONIZED
  CHROMOSPHERIC PLASMAS}

\author{Nicholas A.\ Murphy}
\affil{Harvard-Smithsonian Center for Astrophysics, 60 Garden
  Street, Cambridge, Massachusetts 02138, USA}
\email{namurphy@cfa.harvard.edu}

\author{Vyacheslav S.\ Lukin\footnote{Formerly of U.S. Naval Research Laboratory. Any opinion, findings, and conclusions or recommendations expressed in this material are those of the authors and do not necessarily reflect the views of the National Science Foundation.}}
\affil{National Science Foundation, 4201 Wilson Boulevard, Arlington, Virginia\ 22230, USA}

\keywords{magnetic reconnection --- methods: numerical --- plasmas --- Sun: chromosphere}

\begin{abstract}

Realistic models of magnetic reconnection in the solar chromosphere must take into account that the plasma is partially ionized and that plasma conditions within any two magnetic flux bundles undergoing reconnection may not be the same.  Asymmetric reconnection in the chromosphere may occur when newly emerged flux interacts with pre-existing, overlying flux. We present 2.5D simulations of asymmetric reconnection in weakly ionized, reacting plasmas where the magnetic field strengths, ion and neutral densities, and temperatures are different in each upstream region.  The plasma and neutral components are evolved separately to allow non-equilibrium ionization.  As in previous simulations of chromospheric reconnection, the current sheet thins to the scale of the neutral-ion mean free path and the ion and neutral outflows are strongly coupled. However, the ion and neutral inflows are asymmetrically decoupled. In cases with magnetic asymmetry, a net flow of neutrals through the current sheet from the weak field (high density) upstream region into the strong field upstream region results from a neutral pressure gradient.  Consequently, neutrals dragged along with the outflow are more likely to originate from the weak field region.  The Hall effect leads to the development of a characteristic quadrupole magnetic field modified by asymmetry, but the X-point geometry expected during Hall reconnection does not occur.  All simulations show the development of plasmoids after an initial laminar phase.

\end{abstract}

\maketitle

\section{INTRODUCTION}

Magnetic reconnection in the lower solar atmosphere manifests itself through dynamical events such as jets \citep[e.g.,][]{shibata:2007, Tian:2014:IRIS}, explosive events such as Ellerman bombs \citep[][]{Ellerman:1917, PFChen:2001, CNelson:2013, HYang:2013, Peter:2014:IRIS}, and possibly Type II spicules \citep{depontieu:2007}.  Reconnection events contribute to the heating of the chromosphere \citep[e.g.,][]{jess:2014}, and the contribution of jets and Type II spicules to the mass and energy budgets of the corona and solar wind is under active investigation \citep{depontieu:2011, cranmer:2010, madjarska:2011}.  \cite{arge:1998} proposed that chromospheric reconnection could cause the elemental fractionation that is responsible for the first ionization potential (FIP) effect \citep[see also][]{sturrock:1999A, feldman:2002}.  Partial ionization effects are important not just in the chromosphere but also in molecular clouds, protoplanetary disks, the neutral phases of the interstellar medium, some exoplanetary atmospheres \citep{koskinen:2010}, Earth's ionosphere \citep{leake:2014}, the edge of tokamaks \citep{mladams:thesis}, and dedicated reconnection experiments \citep[e.g.,][]{lawrence:2013}.

Theories and simulations of reconnection have generally assumed that the plasma is fully ionized.  This approximation is valid for the fully ionized solar corona but invalid for weakly ionized plasmas such as the solar chromosphere which has ionization fractions ranging from {$\lesssim$}$0.01$ to {$\sim$}$0.5$.  Partial ionization effects modify the dynamics of reconnection in several ways \citep{zweibel:1989, zweibel:2011}.  Before the onset of reconnection, current sheets thin significantly due to ambipolar diffusion \citep{brandenburg:1994, brandenburg:1995}.  When there is strong coupling between ions and neutrals (e.g., on length scales longer than the neutral-ion mean free path, $\lambda_{ni}$), the effective ion mass is increased by the ratio of the total mass density $\rho$ to the ion mass density $\rho_i$.  This decreases the bulk Alfv\'en speed and consequently the predicted reconnection rate \citep{zweibel:1989}.  The plasma resistivity has contributions from both electron-ion and electron-neutral collisions \cite[e.g.,][]{piddington:1954, cowling:1956, ni:2007}.  

The Hall effect is expected to be important on scales comparable to or less than the ion inertial length $d_i$.  However, the effective ion inertial length is predicted to be enhanced in plasmas with strong ion-neutral coupling: $d_i' = d_i\sqrt{\rho/\rho_i}$ \citep[e.g.,][]{pandey:2008a}.  Consequently, it has been proposed that Hall reconnection may occur at longer length scales or larger densities than would be predicted from the ion density alone \citep{malyshkin:2011, vekstein:2013}.  \citet{malyshkin:2011} predict that enhancement of the Hall effect will lead to fast reconnection in molecular clouds and protoplanetary disks, but that this transition is less likely to be important in the solar chromosphere.  In general, two regimes may be considered.  
When $\lambda_{ni} \ll d_i$, the Hall effect will be enhanced because of strong coupling and fast reconnection is predicted to occur.  When $d_i \lesssim \lambda_{ni} \lesssim d_i'$, some enhancement of the Hall effect due to ion-neutral coupling is also expected. However, as shown previously by \citet[hereafter, \citetalias{leake:partial1}]{leake:partial1} and \citet[hereafter, \citetalias{leake:partial2}]{leake:partial2}, the ion density (and therefore also the local values for $d_i$, $d_i'$, and $\lambda_{ni}$)  may vary by more than an order of magnitude between the reconnection current sheet and the ambient plasma, which makes it difficult to \emph{a priori} evaluate the importance of the Hall effect in determining the rate of reconnection and the structure of the reconnection region in this parameter regime.

The behavior of weakly ionized plasmas can be modeled using either single-fluid or multi-fluid formulations.  A single-fluid approach incorporates the ambipolar diffusion term into the generalized Ohm's law to account for relative drift between ions and neutrals \citep[e.g.,][]{brandenburg:1994, brandenburg:1995}.  This approach is less computationally expensive but implicitly assumes that the ions and neutrals are in ionization equilibrium.  Multi-fluid models evolve each of the ion, neutral, and sometimes electron components separately to allow relative drifts between each of these populations and the fluid to be out of ionization equilibrium \citep[e.g.,][]{meier:2012A}.

In recent years, considerable progress has been made in modeling magnetic reconnection in the lower solar atmosphere. \citet{sakai:2006}, \citet{smith:2008}, \citet{sakai:2008}, and \citet{sakai:2009} performed two-fluid (ion-neutral) simulations of coalescing current loops in the solar chromosphere.  \citet{smith:2008} found that the rate of reconnection in simulations of the upper chromosphere was about $\sim${$20$} times greater than in the lower chromosphere which has a considerably lower ionization fraction.  \cite{sakai:2008} and \citet{sakai:2009} investigated the role of reconnection in penumbra filaments.  These models assume fixed ionization and recombination rates rather than assuming a dependence on temperature and density.  However, it is essential to incorporate rates that depend on physical conditions in order to accurately capture the important role of recombination during chromospheric reconnection.

\citetalias{leake:partial1} and \citetalias{leake:partial2} used the plasma-neutral module of the HiFi framework \citep{lukin:thesis, meier:thesis} to model chromospheric reconnection.  The simulations showed a strong enhancement of ion density inside the current sheet as a result of ions being preferentially dragged along by the reconnecting magnetic field.  Recombination and ion outflow were of comparable importance in the ion continuity equation for removing ions from the current sheet.  The ions and neutral inflows were decoupled, but the ion and neutral outflow jets were strongly coupled.  Late in time, the simulations by \citetalias{leake:partial1} and \citetalias{leake:partial2} showed the development of the secondary tearing instability known as the plasmoid instability \citep[][]{loureiro:2007}.  The increase in the ion and electron densities in the plasmoids led to an increase in the recombination rate that allowed further contraction of the magnetic islands on timescales comparable to the advection time of islands out of the current sheet.  \citet{ni:2015} present one-fluid simulations of the plasmoid instability in the solar chromosphere using the NIRVANA code with and without guide fields.  They find that the onset of the plasmoid instability leads to fast reconnection, and that slow shocks develop near the X-points.  Cases without guide fields show rapid thinning of the current sheet due to ambipolar diffusion and radiative cooling, but these effects are not significant during guide field simulations.

Prior models of magnetic reconnection in weakly ionized plasmas have generally assumed symmetric inflow.  In general, however, there will be some asymmetry in the upstream magnetic field strengths, temperatures, and densities.  The standard model for anenome jets predicts that chromospheric reconnection occurs when newly emerged flux interacts with pre-existing, overlying flux \citep[e.g.,][]{shibata:2007}.  It is reasonable to expect that these different plasma domains will have different plasma parameters, so this configuration naturally leads to asymmetric reconnection.  Because the chromosphere is a dynamic magnetized environment \citep{leenaarts:2007}, asymmetry in the reconnection process is likely to be the norm.

The physics of asymmetric inflow reconnection has been investigated in detail for fully ionized plasmas.  One of the principal applications of this work has been Earth's dayside magnetopause. Asymmetric inflow reconnection has also been investigated in the context of Earth's magnetotail and elsewhere in the magnetosphere \citep{oieroset:2004, muzamil:2014}, the solar atmosphere \citep{NNakamura:2012, murphy:double, YingnaSu:2013:prominence, YangSu:2013}, laboratory experiments \citep{yamada:2007, yoo:2014, murphy:mrx}, and plasma turbulence \citep{servidio:2009, servidio:2010}.  \citet{cassak:asym} performed a scaling analysis for asymmetric inflow reconnection.  They found that the outflow is governed by a hybrid upstream Alfv\'en speed that is a function of the magnetic field strength and density in both upstream regions \citep[see also][]{birn:2010} and that the flow stagnation point in the simulation frame and magnetic field null were not colocated.  The structure and dynamics of asymmetric reconnection in fully ionized collisionless and two-fluid plasmas have been studied previously in several works \citep[e.g.,][]{cassak:hall, cassak:dissipation, malakit:2010, malakit:2013, pritchett:2008:asym, pritchett:2009, mozer:2008, mozer:2009, swisdak:2003, aunai:2013b, aunai:2013a, murphy:mrx}. Simulations of the plasmoid instability during reconnection with asymmetric upstream magnetic fields by \citet{murphy:plasmoidasym} showed that the resultant magnetic islands developed primarily into the weak field upstream region.  Because the reconnection jets impacted the islands obliquely rather than directly, the islands developed net vorticity.  In addition to asymmetric inflow reconnection, several groups have investigated asymmetric outflow reconnection \citep[e.g.,][]{oka:2008, murphy:asym, murphy:retreat} and reconnection with three-dimensional asymmetry \citep[e.g.,][]{AlHachami:2010, wyper:2013}.  

Observational investigations of partially ionized reconnection in the chromosphere are challenging because the dissipation length scales are significantly shorter than can be resolved with current instrumentation.  Diagnosing the chromospheric magnetic field is an important but very difficult problem that requires inversion of spectropolarimetric data \citep[e.g.,][]{kleint:2012}, and this will be even more challenging for short-lived dynamical events.  Comparisons between simulations and observations should therefore concentrate on the large-scale consequences of chromospheric reconnection which depend on small-scale processes.  A complementary approach to solar observations is to study partially ionized reconnection in the laboratory.  \citet{lawrence:2013} have performed such studies at the Magnetic Reconnection Experiment \citep[MRX;][]{yamada:1997a}.

Our motivation is to investigate the role of asymmetry on the small-scale physics of reconnection in partially ionized chromospheric plasmas.  While there are similarities to symmetric cases, there are also physical effects such as neutral flows through the current sheet that do not occur in cases with symmetric inflow.  We use the plasma-neutral module of the HiFi modeling framework \citep{lukin:thesis} to perform simulations of magnetic reconnection with asymmetric upstream magnetic field strengths, temperatures, and/or densities in the weakly ionized solar chromosphere.  In Section \ref{numerical}, we describe the numerical method and problem setup (see also \citetalias{leake:partial1} and \citetalias{leake:partial2}). In Section \ref{global}, we describe the global dynamics of reconnection, the structure of the current sheet, the role of the Hall effect, the dynamics of the plasmoid instability, the motion of the X-point, and ion/neutral flows at the X-point.  We discuss our results in Section \ref{discussion}.

\section{NUMERICAL MODEL AND PROBLEM SETUP\label{numerical}}

The HiFi framework\footnote{See \begin{tt}http://faculty.washington.edu/vlukin/HiFi\_Framework.html\end{tt}}
\citep{lukin:thesis, glasser:2004} uses a spectral element spatial representation and implicit time advance to solve systems of partial differential equations.  The modular approach makes new physical models straightforward to implement \citep[e.g.,][]{gray:2010B, lukin:2011, ohia:2012, le:2013, stanier:2013, browning:2014, elee:2014}.  In this paper, we use the module for partially ionized, reacting plasmas \citep[][\citetalias{leake:partial1}, \citetalias{leake:partial2}]{meier:thesis}.  These plasmas consist of neutral and ionized hydrogen and electrons.  A multi-fluid approach allows the plasma and neutral components to be modeled separately.  We summarize the equations used for our 2.5D simulations in this section, but direct the reader to \citetalias{leake:partial1} and \citetalias{leake:partial2} for further detail.  In general, we use the notation and conventions from \citetalias{leake:partial1} and \citetalias{leake:partial2}\@.  The subscripts `n', `i', and `e' refer to neutrals, ions, and electrons, respectively.  

\subsection{Normalizations\label{normalizations}}

Following \citetalias{leake:partial1} and \citetalias{leake:partial2}, we normalize the fluid equations to values characteristic of the lower chromosphere.  We choose a characteristic length scale of $L_\star\equiv 1\times 10^4$ m, a characteristic number density of $n_\star \equiv 3\times 10^{16}$ m$^{-3}$, and a characteristic magnetic field strength of $B_\star \equiv 1\times 10^{-3}$ T\@.  From these quantities, we derive additional normalizing values to be $V_\star\equiv B_\star / \sqrt{\mu_0 m_p n_\star} = 1.26\times 10^5$ m\,s$^{-1}$ for velocity, $t_\star \equiv L_\star / V_\star = 0.0794$ s for time, $T_\star \equiv B_\star^2 / k_B \mu_0 n_\star = 1.92 \times 10^6$ K for temperature, $P_\star \equiv B_\star^2/\mu_0 = 0.796$ Pa for pressure, $J_\star \equiv B_\star/\mu_0 L_\star = 7.96\times 10^{-2}$ A\,$\cdot$\,m$^{-2}$ for current density, and $\eta_\star \equiv \mu_0 L_\star V_\star = 1.58 \times 10^3$ $\Omega$\,{$\cdot$}\,m for resistivity.  Unless otherwise indicated (e.g., Sections \ref{irce} and \ref{initial}), the equations presented for the simulation will be in dimensionless units according to these normalizations.  

\subsection{Ionization and Recombination\label{irce}}

The HiFi module for partially ionized plasmas includes both ionization and recombination of hydrogen to allow departures from ionization equilibrium \citepalias{leake:partial1, leake:partial2}.  The ionization and recombination rates are given by
\begin{eqnarray}
  \Gamma^{ion}_n & \equiv & -n_n \nu^{ion}, \label{ionrate}  \\
  \Gamma^{rec}_i & \equiv & -n_i \nu^{rec}, \label{recrate}
\end{eqnarray}
with $\Gamma^{ion}_i = -\Gamma^{ion}_n$ and $\Gamma^{rec}_i =
-\Gamma^{rec}_n$.  The ionization frequency of hydrogen is approximated to be
\begin{equation}
  \nu^{ion} = \frac{n_e A}{X + \phi_{ion}/T_e^*}
  \left(\frac{\phi_{ion}}{T_e^*}\right)^K 
  \exp\left(-\frac{\phi_{ion}}{T_e^*}\right)
  \mbox{~m}^3\mbox{\,s}^{-1}, \label{ionfreq}
\end{equation}
with $A = 2.91\times 10^{-14}$, $K=0.39$, $X=0.232$, and $\phi_{ion} = 13.6$ eV \citep{voronov:1997}. Here, $T_e^*$ is the electron temperature in eV\@. \citet{smirnov:2003} approximates the recombination frequency of hydrogen to be
\begin{equation}
  \nu^{rec} = 2.6 \times 10^{-19} \frac{n_e}{\sqrt{T_e^*}} 
    \mbox{~m}^3\mbox{\,s}^{-1}
    \label{recfreq}.
\end{equation}
At a characteristic temperature of $9000$ K, this corresponds to an equilibrium ionization fraction of $4.1\times 10^{-4}$.  These expressions do not include the consequences of non-local thermodynamic equilibrium radiative transfer.  For example, if Lyman $\alpha$ is optically thick, then there will be more neutral hydrogen in the $n=2$ state than predicted from these expressions which would allow for enhanced ionization. At lower temperatures, ionization from low-FIP elements may contribute more to the electron density than hydrogen.

\subsection{Multi-fluid Equations\label{fluidequations}}

In this section, we summarize the equations solved by the plasma-neutral module of HiFi.  The equations are described more thoroughly by \citetalias{leake:partial1} and \citetalias{leake:partial2} \citep[see also][]{meier:thesis, meier:2012A}.  Our simulations include the modifications and extensions to the model described by \citetalias{leake:partial2}.  

The ion and neutral continuity equations are
\begin{eqnarray}
  \frac{\partial n_i}{\partial t} + \nabla\cdot
  \left(n_i\vi\right) & = & \Gamma^{rec}_i + \Gamma^{ion}_i,
  \\ 
  \frac{\partial n_n}{\partial t} + \nabla\cdot
  \left(n_n\vn\right) & = & \Gamma^{rec}_n + \Gamma^{ion}_n.
\end{eqnarray}
These equations include ionization and radiative recombination as described in Section \ref{irce}, \citetalias{leake:partial1}, and \citetalias{leake:partial2}.  We assume quasineutrality such that $n_i=n_e$.

The ion and neutral momentum equations are given by
\begin{eqnarray}
  \frac{\partial }{\partial t}
  \left(m_in_i\vi\right) + 
  \nabla \cdot \left( m_in_i\vi\vi + \ptensi + \ptense \right) = 
  \hspace{6cm}\nonumber\\ \hspace{2cm}
  \J \times \B + \Rini + \Gamma^{ion}_im_i\vn - \Gamma^{rec}_nm_i\vi 
  + \Gamma^{cx}m_i\left(\vn-\vi\right) +\Rcxin - \Rcxni, \label{momentum_plasma}
  \\
  \frac{\partial }{\partial t}
  \left(m_in_n\vn\right) + 
  \nabla \cdot \left( m_in_n\vn\vn + \ptensn \right) = 
  \hspace{6cm}\nonumber\\ \hspace{2cm}
  -\Rini + \Gamma^{rec}_nm_i\vi - \Gamma^{ion}_im_i\vn
  + \Gamma^{cx} m_i\left(\vi-\vn\right)
  -\Rcxin + \Rcxni 
  . \label{momentum_neutrals}
\end{eqnarray}
The Lorentz force acts directly on the plasma but not the neutrals, while the neutral pressure gradient acts directly on the neutrals but not the plasma.  Coupling between the plasma and neutrals is achieved though many of the terms on the right hand side of Eqs.\ \ref{momentum_plasma} and \ref{momentum_neutrals}. The term $\Rini$ represents momentum transfer from neutrals to ions due to identity preserving collisions and is given by
\begin{equation}
    \Rini = m_{in} n_i \nu_{in} \left(\vn-\vi\right), \label{rinidef}
\end{equation}
where $m_{in} = m_i m_n /(m_i+m_n)$ and the collision frequency $\nu_{in}$ is given by Eq.\ 7 of \citetalias{leake:partial2}.  The pressure tensor for species $\alpha$ is
\begin{equation}
    \mathsf{P}_\alpha=P_\alpha\mathsf{I} + \pi_\alpha
    \label{ptensdef},
\end{equation}
where $P_\alpha$ is the scalar pressure and $\mathsf{I}$ is the identity tensor.  The viscous stress tensor is then
\begin{equation}
    \pi_\alpha = - \xi_\alpha\left[\nabla\mathbf{V}_\alpha + \left(\nabla\mathbf{V}_\alpha\right)^\top\right],
\end{equation}
with $\xi_\alpha$ as the isotropic dynamic viscosity coefficient.  The terms that depend on the ionization and recombination rates represent momentum transfer by particles that change identity from neutrals to ions and vice-versa. We follow \citetalias{leake:partial2} and include charge exchange. Here, $\Gamma^{cx}$ is the charge exchange reaction rate and $\mathbf{R}^{cx}_{\alpha\beta}$ is the momentum transfer due to charge exchange rections from species $\beta$ to species $\alpha$.  Charge exchange leads to increased momentum and energy transfer between species by about a factor of two, and therefore more effective coupling.  We use the formulation for charge exchange given by Eqs.\ 4--6 of \citetalias{leake:partial2} \citep[see also][]{meier:thesis, meier:2012A, leake:partial1, Barnett:1990}.

We adjust the collisional cross sections to be $\Sigma_{in}=\Sigma_{ni}=5\times 10^{-19}$ m$^2$ (used in Eq.\ 7 of \citetalias{leake:partial2} to calculate $\nu_{in}$) and $\Sigma_{nn}=5\times 10^{-19}$ m$^2$ (used in Eq.\ 9 of \citetalias{leake:partial2} to calculate $\nu_{nn}$).  These values differ from \citetalias{leake:partial1} and \citetalias{leake:partial2}, but the value for $\Sigma_{in}$ is consistent with \citet{khomenko:2012} and \citet{ni:2015}.  These cross sections are a function of energy, but we use constant values appropriate for the chromosphere as a simplifying assumption \citep[see also][]{draine:1983}.  The neutral and ion viscosity coefficients $\xi_n$ and $\xi_i$ are set by Eq.\ 8 of \citetalias{leake:partial2} using the revised value of $\Sigma_{nn}$.  While $\xi_n$ is a function of local physical conditions, $\xi_i$ and $\xi_e$ are constant throughout the domain and correspond to the parallel component of the Braginskii viscosity for each species computed using the mean of the asymptotic upstream densities and temperatures.  We use $\xi_i=2.8\times 10^{-6}$ and $\xi_e=4.6\times 10^{-14}$ with a normalization of $\xi_\star=m_p n_\star L_\star V_\star$.  

The neutral-ion mean free path
\begin{equation}
    \lambda_{ni} = \frac{V_{T,n}}{\nu_{ni}^\dagger}  \label{lambdanidef}
\end{equation}
governs the lengths scales above which the neutrals and ions are coupled to each other.  Here, the neutral thermal velocity is $V_{T,n}=\sqrt{2 k_B T_n/m_n}$ where $k_B$ is the Boltzmann constant, the neutral temperature is $T_n$, and the neutral mass is $m_n$.  The frequency $\nu_{ni}^\dagger = \nu_{ni} + \nu_{ni}^{CX}$ includes contributions from the neutral-ion collision frequency $\nu_{ni}$ (see Eq.\ 16 of \citetalias{leake:partial1}) and the neutral-ion charge exchange frequency $\nu_{ni}^{CX}$ (see also Eqs.\ 4--6 of \citetalias{leake:partial2}).  The ions and neutrals will be coupled on length scales much longer than $\lambda_{ni}$, and decoupled on scales much shorter than $\lambda_{ni}$.

The energy equations for the plasma and neutral components are given by Eqs.\ 19 and 20 of \citetalias{leake:partial1}.  Both equations include frictional heating due to identity preserving collisions, thermal transfer due to changes in identity, and the effects of charge exchange.  The energy equation for the plasma component combines the electron and ion energy equations and includes Ohmic heating, optically thin radiative losses, and anisotropic thermal conduction \citepalias[Eq.\ 21 of][]{leake:partial1}. The neutral energy equation includes isotropic thermal conduction that depends on plasma parameters and is set by Eq.\ 10 of \citetalias{leake:partial2}.  Neutral thermal conduction dominates thermal diffusion, in part due to rapid thermal transfer between neutrals and ions.  For characteristic values of $T=9000$ K and $n_n=7.6\times 10^{18}$ m$^{-3}$, the neutral thermal conductivity is $\kappa_n=3.05\times 10^{22}$ m$^{-1}$\,s$^{-1}$. We assume that the ion and electron temperatures are equal, but the neutral temperature is evolved separately.  

The generalized Ohm's law for this paper is given by
\begin{equation}
    \E + \vi\times\B 
    =
    \eta\J
    + \frac{\J\times\B}{e n_e} 
    - \frac{\nabla P_e}{e n_e}
    - \frac{m_e \nu_{en}}{e} \left(\vi-\vn\right) \label{genohms}
\end{equation}
The resistivity includes both electron-ion and electron-neutral collisions and is given by
\begin{equation}
    \eta = \frac{m_e n_e \left(\nu_{ei} + \nu_{en}\right)}{\left( e n_e \right)^2},
\end{equation}
where the electron-ion collision frequency $\nu_{ei}$ and the electron-neutral collision frequency $\nu_{en}$ are functions of number density and temperature and are given by Eq.\ 13 of \citetalias{leake:partial2} with $\Sigma_{en}=\Sigma_{ne}=1\times 10^{-19}$ m$^{2}$ as used previously in \citetalias{leake:partial1}, \citetalias{leake:partial2}, \citet{khomenko:2012}, and \citet{ni:2015}.  The resistivity therefore depends on plasma parameters and is not a constant as in \citetalias{leake:partial1}.  We include the Hall term and evolve scalar plasma and neutral pressures. 

\subsection{Initial Conditions\label{initial}}

Here we describe the procedure to establish an approximate initial equilibrium with asymmetric upstream magnetic field strengths, densities, temperatures, and ionization fractions.  Before proceeding, we define several parameters to quantify asymmetries in different fields.  The magnetic, temperature, number density (of ions and neutrals), and ionization fraction asymmetries are defined as
\begin{eqnarray}
  \basym & \equiv & \frac{B_{2}}{B_{1}}, \label{basym}
  \\  
  \tasym & \equiv & \frac{T_{2}}{T_{1}}, \label{tasym}
  \\
  \nasym & \equiv &
  \frac{n_{i,2}+n_{n,2}}{n_{i,1}+n_{n,1}}, \label{nasym}
  \\
  \fasym & \equiv & \frac{f_2}{f_1}, \label{fasym}
\end{eqnarray}
where the subscripts `1' and `2' correspond to the asymptotic upstream magnitudes of each field for $y>0$ and $y<0$, respectively, and the ionization fraction is defined as
\begin{equation}
  f \equiv \frac{n_i}{n_i + n_n}. \label{ionfrac}
\end{equation}
All of these quantities are functions of time, so the subscript `0' in expressions below indicates correspondence to the initial conditions.

The equilibrium magnetic field is specified as a modified Harris sheet,
\begin{equation}
  B_{x0}\left(y\right) = B_{1,0} \left[ \frac{\tanh \left(
      \frac{y}{\thick} - b\right) + b}{1+b}
    \right],  \label{initial_b}
\end{equation}
where $\thick$ is the initial thickness of the current sheet \citep[see also][]{birn:2008, birn:2010, murphy:double, murphy:plasmoidasym}.  The initial magnetic asymmetry is given by $\basym_0 = (1-b)/(1+b)$ with the convention that $0 \leq b < 1$ so that $B_{2,0}\leq B_{1,0}$.  We describe the in-plane magnetic field using the magnetic flux, $A_z$, such that 
\begin{equation}
  \mathbf{B} = \nabla\times \left(A_z\zhat\right) + B_z\zhat. \label{fluxdef}
\end{equation}
The flux corresponding to Eq.\ \ref{initial_b} is 
\begin{equation}
  A_{z0}(y) = \frac{B_{1,0}}{1+b}
           \left[
             \thick
             \ln\cosh\left(\frac{y}{\thick}-b\right)
             + b y
           \right].
\end{equation}
The out-of-plane current density is then
\begin{equation}
  J_{z0}(y) = - \frac{B_{1,0}}{\mu_0\thick}
  \left[
  \frac{
    \mathrm{sech}^2{\left(\frac{y}{\thick} - b
      \right)}}{1+b} \right]. \label{initial_j}
\end{equation}

The initial temperature is set using the relations
\begin{eqnarray}
  T_0(y) & = & T_{1,0} 
  \left[ 1 + 
  \left( \tasym_0 - 1 \right)
  \left( 1 - \zeta^2   \right)
  \right], \label{tempinit}
  \\
  \zeta &=& \frac{1}{2} 
  \left[ 1 + \tanh
     \left(
         \frac{y}{\thick} - b
     \right) 
  \right] . \label{zetadef}
\end{eqnarray}
The initial ionization fraction is found numerically by equating the ionization rate $\Gamma^{ion}_i$ with the recombination rate $\Gamma^{rec}_i$ using Eqs.\ \ref{ionrate}--\ref{recfreq}.  The initial conditions are therefore in ionization equilibrium.  If the initial temperature is non-uniform, then the initial ionization fraction will also be non-uniform: if $\tasym_0 \ne 1$, then $\fasym_0 \ne 1$. This is in contrast to \citetalias{leake:partial1} and \citetalias{leake:partial2} which both assume that the temperature and ionization fraction are both initially uniform.

We define $\ptot$ to be the sum of the plasma and neutral pressures and the plasma pressure $P_p$ to be the sum of the electron an ion pressures,
\begin{eqnarray}
  \ptot & \equiv & P_n + P_p,
  \\
  P_p  & \equiv & P_i + P_e. 
\end{eqnarray}
For equal temperatures, the neutral and plasma pressures are related to the ionization fraction by
\begin{eqnarray}
  P_n &=& \ptot \left(\frac{1 - f}{1 + f}\right),
  \\
  P_p &=& \ptot \left(\frac{ 2 f }{1 + f}\right),
\end{eqnarray}
where we recall that the total number of particles depends on the ionization fraction and that electrons are included in the plasma pressure.  We calculate ${\ptot}_{,0}$ to balance the Lorentz force associated with the magnetic field profile given by Eq.\ \ref{initial_b},
\begin{equation}
  P_{\mathrm{tot},0}(y) = \left( 1 + \beta_{1,0} \right)
  \frac{B_{1,0}^2}{2\mu_0} - \frac{B_{x0}(y)^2}{2\mu_0}, \label{ptot}
\end{equation}
where $\beta_{1,0}$ is the ratio of ${\ptot}_{,0}$ to the magnetic pressure for $y \gg 0$.  The number densities are then given by
\begin{eqnarray}
  n_n &=& \frac{P_n}{k_B T},
  \\ 
  n_i &=& \frac{P_p}{2 k_B T}.
\end{eqnarray} 
We assume that the ions and electrons have equal temperatures and number densities.

\begin{figure}[t!]
  \begin{center}
    \includegraphics{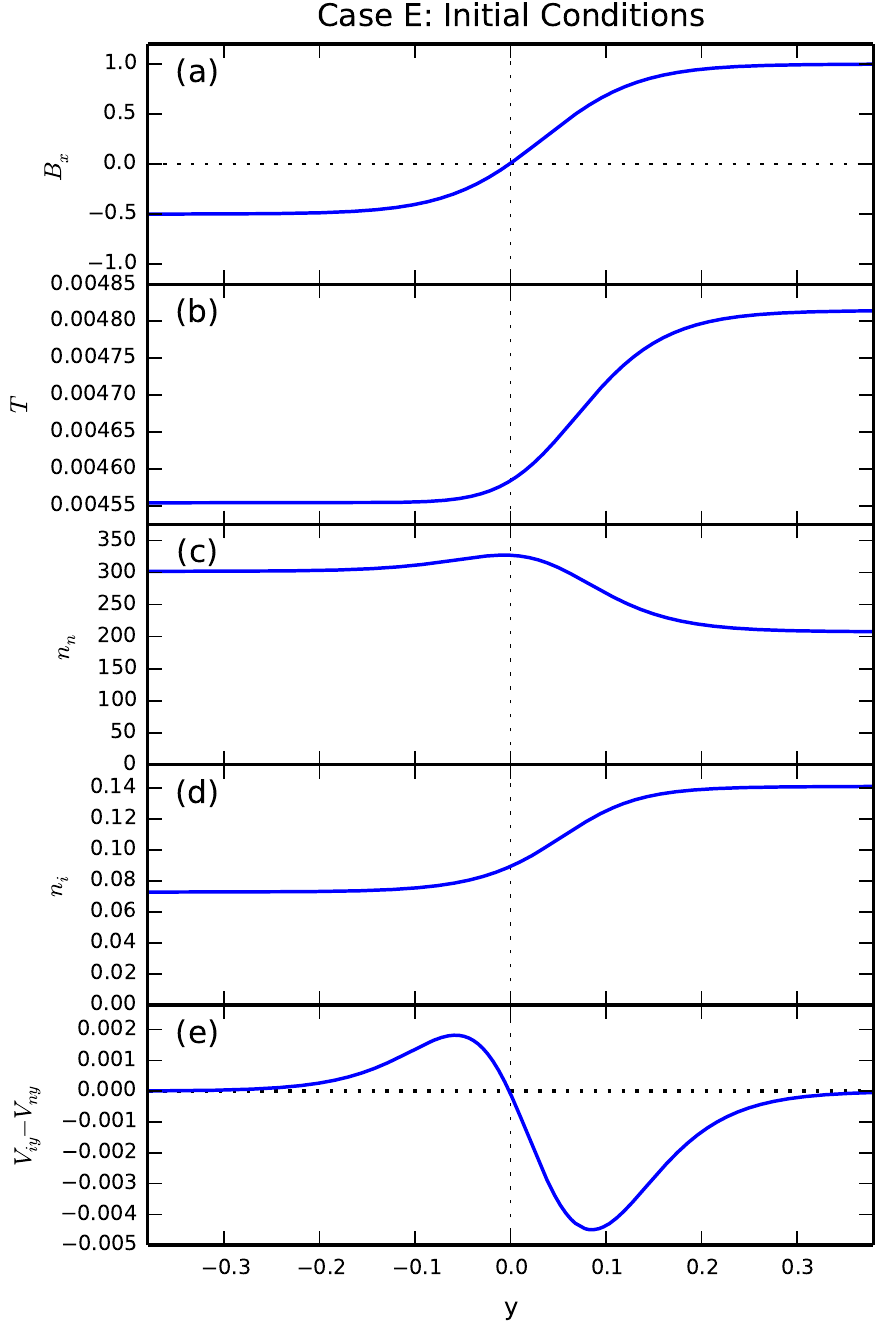}
  \end{center}
  \caption{The initial conditions for case E with $\basym_0=1$, $\tasym_0=0.95$, $\nasym_0=1.47$, and $\fasym_0=0.35$.
    \label{ICplot}}
\end{figure}

The above magnetic and pressure profiles satisfy the relation
\begin{equation}
  0 = - \nabla \left( P_p + P_n \right) + \J \times \B.
\end{equation}
However, there must also be a relative velocity between the plasma and neutrals to allow coupling via the momentum equations of the different species through collisions and charge exchange.  In the absence of charge exchange, the steady state momentum equations for the plasma and neutral components are
\begin{eqnarray}
  0 &=& -\nabla P_p + \J\times\B + \Rini,
  \\
  0 &=& -\nabla P_n - \Rini,
\end{eqnarray}
respectively, where the frictional force $\Rini$ is defined in Eq.\ \ref{rinidef}. In this approximation, the initial ion and neutral velocities along the inflow direction can be chosen so that the initial configuration contains no net force on either of the plasma and neutral components,
\begin{eqnarray}
V_{iy0}-V_{ny0} = \frac{\partial P_n/\partial y}{m_{in}n_i\nu_{in}}, \label{viy0def}
\end{eqnarray}
where the neutral pressure gradient is calculated analytically using the expression
\begin{equation}
  \frac{\partial P_n}{\partial y} = -
  \left( \frac{1-f_0}{1+f_0} \right)
  \left( \frac{B_{1,0}}{\mu_0 \lambda_\psi \left(1+b\right)} \right)
  B_{x0}(y)\,
  \mathrm{sech}^2\left( \frac{y}{\lambda_\psi} - b\right) .
\end{equation}
Because our simulations include charge exchange, momentum transfer between the ions and neutrals is of order twice as effective as the case with frictional coupling alone.  In lieu of an exact solution to the steady state momentum equation, we reduce the initial ion-neutral drift velocity given in Eq.\ \ref{viy0def} by a factor of two so that the initial conditions for both ions and neutrals are in approximate but not exact force balance.  We set $V_{iy0}$ and $V_{ny0}$ so that they have the same magnitude but opposite sign.  Example initial conditions are shown in Fig.\ \ref{ICplot}\@.

The initial state does not represent an exact equilibrium, which leads to an outward moving pulse along the inflow direction that propagates from the initial current sheet.  This pulse is mostly damped by placing a layer of significantly enhanced viscosity near the conducting walls at $y=\pm 1$.  

\subsection{Boundary Conditions\label{BCs}}

We simulate a half-domain that extends from $0\leq x\leq L_x$ and from $-L_y \leq y \leq L_y$, where $x$ is the outflow direction and $y$ is the inflow direction.  We assume that the fields $A_z$, $n_i$, $n_n$, $p_i$, $p_n$, $V_{iy}$, $V_{ny}$, $V_{iz}$, $V_{nz}$, and $J_z$ are symmetric about $x=0$ [e.g., $A_z(x,y)=A_z(-x,y)$], and that the fields $V_x$ and $B_z$ are antisymmetric about $x=0$ [e.g., $V_x(x,y)=-V_x(-x,y)$].  We assume that all fields are periodic along the $x$ direction with a period of $2L_x$.  We apply perfectly conducting, zero-flux boundary conditions along the boundaries at $y=\pm L_y$.  

The choice of initial and boundary conditions enforces that the reconnection outflow is symmetric about $x=0$.  The development of the plasmoid instability under these conditions often leads to the formation of a magnetic island centered about $x=0$ that cannot be advected out of the system.  In general, there will be some degree of asymmetry along the outflow direction that can modify the internal structure of the reconnection layer and the dynamics of reconnection \citep{oka:2008, murphy:asym, murphy:retreat}.  This can be achieved by simulating the whole domain of interest and allowing the driving term, the outflow boundaries, or the initial perturbations to be asymmetric \citep[e.g.,][]{murphy:plasmoidasym}.  

\subsection{Initializing Reconnection\label{initialization}}

To initialize reconnection, we apply a source function to the evolution equation for magnetic flux of the form
\begin{equation}
    \frac{\dif A_z}{\dif t} = 
    \epsilon \lambda_\psi 
    \Lambda\left(x,y\right)
    \Omega\left(x,y\right)
    \Theta\left(t\right)
\end{equation}
where
\begin{eqnarray}
\Lambda\left(x,y\right) &=&
    \exp\left[
        - \left( \frac{x}{h_x} \right)^2
        - \left( \frac{y}{h_y} \right)^2
    \right]
    \left\{ 1 - \frac{1}{2} \exp
        \left[
            - 3 \left( \frac{y}{h_y} \right)^2
        \right] 
    \right\},
    \\
    \Omega\left(x,y\right) &=&
        \left[ 1 - \left( \frac{x}{L_x} \right)^{16} \right]
        \left[ 1 - \left( \frac{y}{L_y} \right)^{16} \right],
    \\  
    \Theta\left(t\right) &=&
    \frac{1}{2}
    \left[
        1 - \cos\left( \frac{2\pi t}{t_{D}}\right)
    \right].
\end{eqnarray}
The spatial length scales for this source function are given by $h_x = 4\lambda_\psi$ and $h_y = \lambda_\psi$.  The function $\Lambda(x,y)$ localizes the source function in a form akin to a tearing mode eigenfunction to create an X-point at the field reversal along $x=0$.  The function $\Omega(x,y)$ ensures that the pulse goes to zero along the outer boundaries.  The pulse is applied between $0\leq t \leq t_{D}$ with a waveform given by $\Theta\left(t\right)$.  We use $\epsilon=0.05$ and $t_{D}=5$.  The application of this electric field ceases before the reconnection layer has had a chance to develop.  In contrast, \citetalias{leake:partial1} applied a small, localized perturbation to the magnetic flux while \citetalias{leake:partial2} applied a small amplitude localized rotational flow perturbation to initialize reconnection.

\section{LOCAL AND GLOBAL DYNAMICS OF RECONNECTION\label{global}}

\begin{deluxetable}{ccccccccccccc}
\tablecolumns{7}
\tablewidth{0in}
\tablecaption{Simulation initial parameters\label{table_simpars}}
\tablehead{
\colhead{Case} &
\colhead{$B_{1,0}$} &
\colhead{$B_{2,0}$} & 
\colhead{$T_{1,0}$} &
\colhead{$T_{2,0}$} &
\colhead{$n_{1,0}$} &
\colhead{$n_{2,0}$} &
\colhead{$f_{1,0}$} & 
\colhead{$f_{2,0}$} &
\colhead{$\beta_{1,0}$} &
\colhead{$\beta_{2,0}$} & 
\colhead{$\lambda_{ni,1,0}$} &
\colhead{$\lambda_{ni,2,0}$}
}
\startdata
A & $7.9$ & $7.9$ & $9000$ & $9000$ & $7.6$ & $7.6$ & $0.00041$ & $0.00041$ & $3.8$ & $3.8$ & $220$ & $220$ \\
B & $7.9$ & $7.9$ & $8750$ & $9250$ & $7.8$ & $7.4$ & $0.00024$ & $0.00068$ & $3.8$ & $3.8$ & $366$ & $136$ \\
C & $10$ & $5$    & $9000$ & $9000$ & $8.8$ & $6.4$ & $0.00041$ & $0.00041$ & $11.0$ & $2.0$ & $261$ & $190$ \\
D & $10$ & $5$    & $8750$ & $9250$ & $6.6$ & $8.6$ & $0.00024$ & $0.00068$ & $2.0$  & $11.0$ & $432$ &$117$ \\
E & $10$ & $5$    & $9250$ & $8750$ & $6.2$ & $9.1$ & $0.00068$ & $0.00024$ & $2.0$ & $11.0$& $163$ & $314$ \\
\enddata
\vspace{-0.8cm}
\tablecomments{The units are G for magnetic field, K for temperature, $10^{18}$ m$^{-3}$ for number density (including both neutrals and ions), and m for the neutral-ion mean free paths. The neutral-ion mean free path $\lambda_{ni}$ is calculated using Eq.\ \ref{lambdanidef} and includes charge exchange reactions.  When calculating the charge exchange cross section using Eqs.\ 5--6 of \citetalias{leake:partial2}, we use that the relative velocity between ions and neutrals is much less than the neutral and ion thermal speeds for the initial conditions.
}
\end{deluxetable}

\afterpage{\clearpage}

We present five simulations to investigate the impact of asymmetry during magnetic reconnection in partially ionized chromospheric plasmas.  The simulation parameters are shown in Table \ref{table_simpars}.  Case A is the symmetric test case.  Case B has asymmetric upstream temperatures, and consequently asymmetric upstream densities and ionization fractions, but symmetric upstream magnetic field strengths.  Case C has symmetric upstream temperatures but a factor of two difference in the upstream magnetic field strengths.  Cases D and E have asymmetric temperatures and magnetic field strengths.  We keep three parameters constant between runs: the sum of the initial magnetic pressure, neutral pressure, and plasma pressure which equals $1.495$ ($1.19$ Pa); the mean of the initial asymptotic upstream magnetic energy densities which is given by  $\frac{1}{2}\left( B_{1,0}^2/{2} + B_{2,0}^2/{2}\right) = 0.3125$ ($0.249$ J\,m$^{-3}$); and the mean of the initial asymptotic upstream temperatures which is $(T_{1,0}+T_{2,0})/2 = 4.69\times 10^{-3}$ ($9000$ K).  

The domain size for all simulations is $(L_x,L_y)=(2,1)$.  The resolution in Case A is $m_x=256$ elements along the outflow direction and $m_y=128$ elements along the inflow direction.  The resolution in Cases B--E is $m_x=m_y=256$.  We use sixth order basis functions for all simulations, resulting in effective total resolution of $(M_x,M_y)=6(m_x,m_y)$.  Grid packing is used to concentrate mesh in the reconnection region.  In Case A, the current sheet does not move from $y=0$ so the mesh packing along the inflow direction is concentrated to a thin region near $y=0$.  In Cases B--D, the current sheet drifts slowly in the $-\yhat$ direction so the highest resolution is needed over a much longer distance to resolve the dynamics \citep[see also][]{murphy:double, murphy:plasmoidasym}.  High resolution along the outflow direction helps capture the front end of the reconnection jet as well as the dynamics of the plasmoid instability.  The resolution along the outflow direction from $0\leq x \lesssim 1.2$ and $1.95\lesssim x \leq 2$ is approximately twice as high as from $1.2\lesssim x \lesssim 1.95$.  

\subsection{Structure of Reconnection Region\label{structure}}

\begin{figure}[tp]
    \begin{center}
    \vspace{-7mm}
    \includegraphics[height=7.5in]{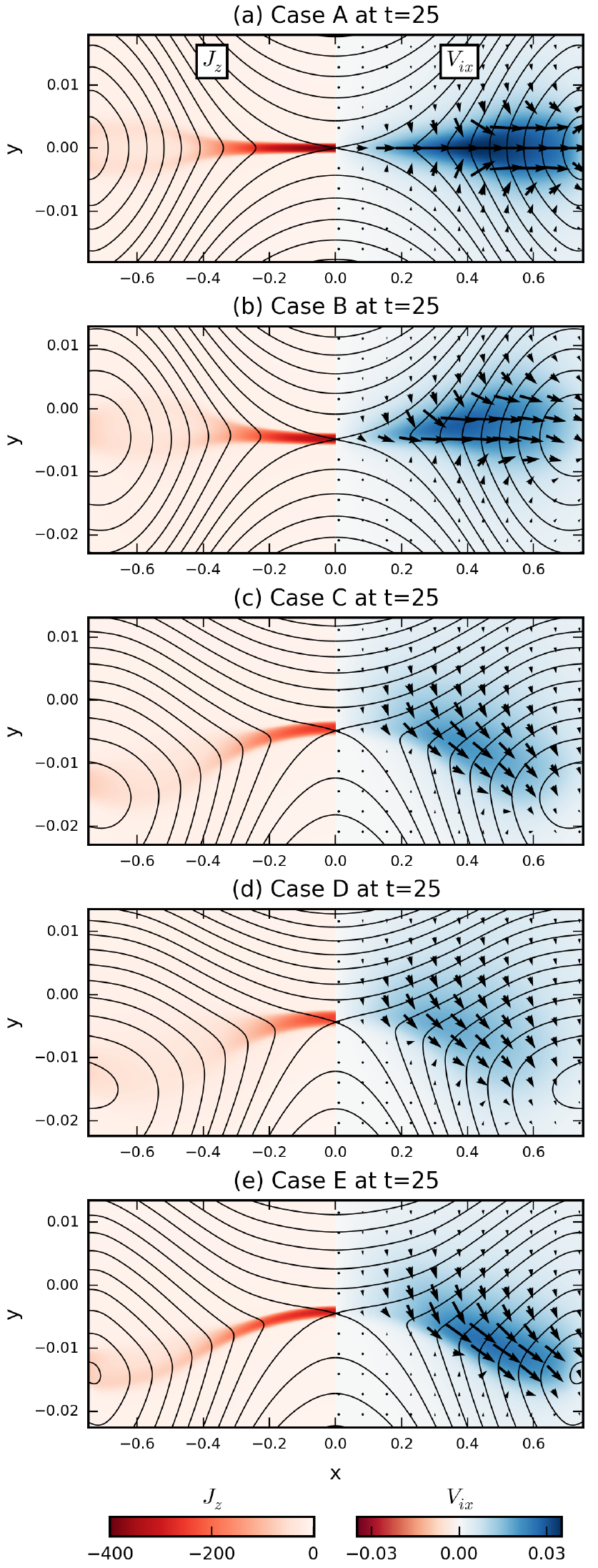}
    \end{center}
    \vspace{-7mm}
    \caption{The out-of-plane current density $J_z$ (left) and the ion outflow (right) at $t=25$ for Cases A--E\@.  The right half of each plot shows color contours of the outflow component of ion velocity $V_{ix}$, ion velocity vectors, and contours of magnetic flux $A_z$.  This image is scaled significantly along the $y$ direction which exaggerates the inflow $y$-component of velocity.
        \label{jzvxplot}}
\end{figure}

\begin{figure}[tp]
    \begin{center}
    \vspace{-7mm}
    \includegraphics[height=7.5in]{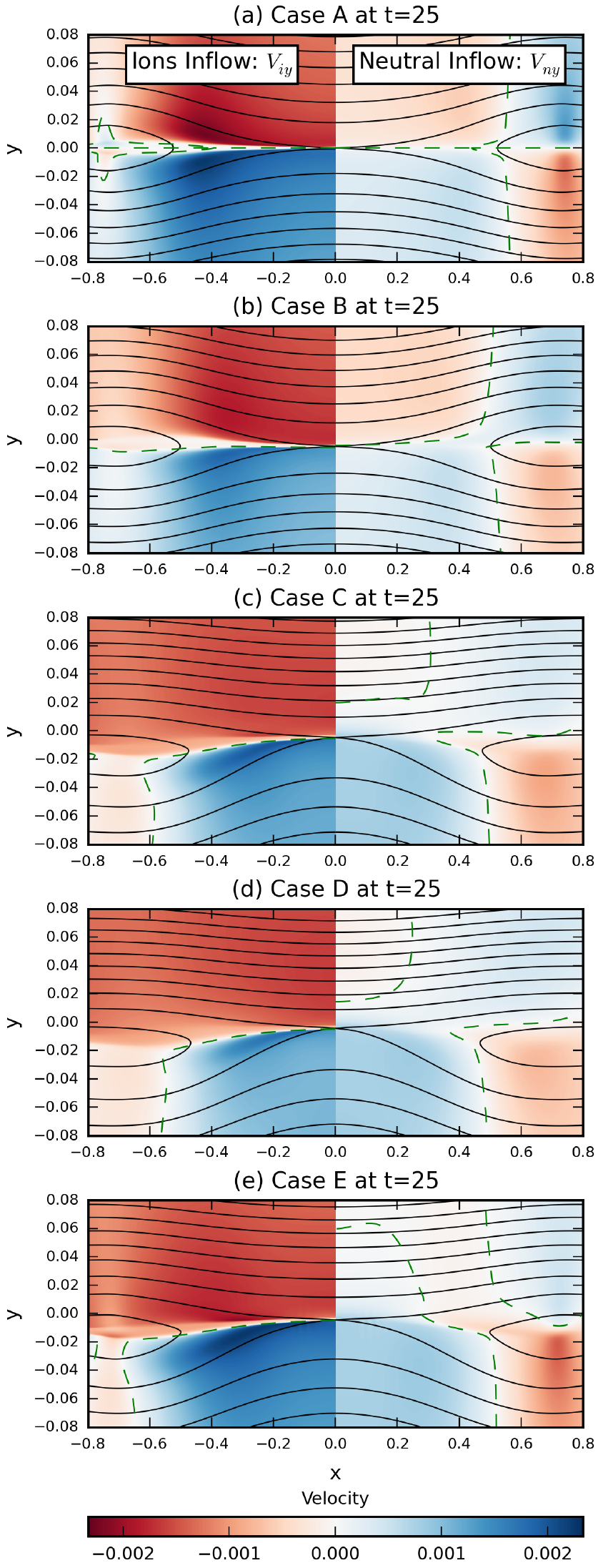}
    \end{center}
    \vspace{-7mm}
    \caption{The inflow components of the ion velocity $V_{iy}$ (left) and the neutral velocity $V_{ny}$ (right) in the reconnection region at $t=25$ for Cases A--E\@.  The solid black contours represent the magnetic flux $A_z$.  The dashed green contour indicates the locations where the inflow component of velocity equals zero.  
    \label{flowplot}}
\end{figure}

\begin{figure}[tp]
    \begin{center}
    \vspace{-7mm}
    \includegraphics[height=7.5in]{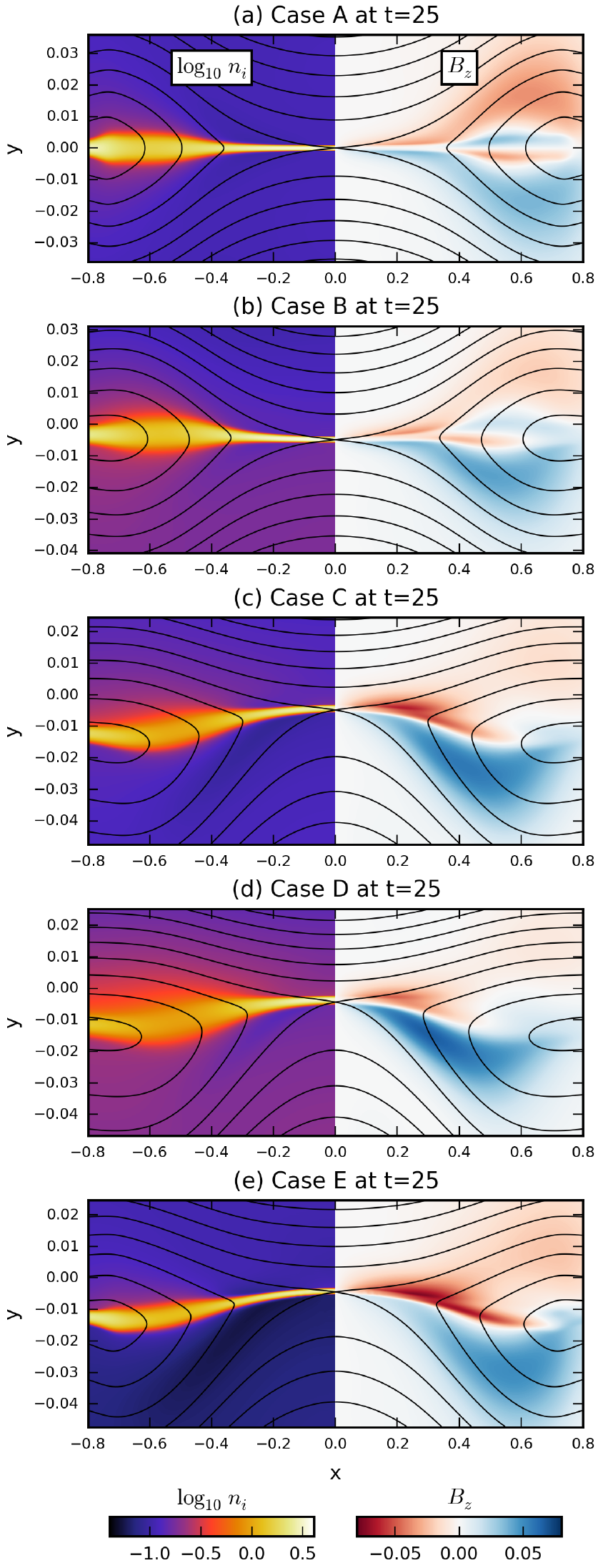}
    \end{center}
    \vspace{-7mm}
    \caption{The logarithm of the ion number density $\log_{10}\,{n_i}$ (left) and the out-of-plane magnetic field $B_z$ (right) at $t=25$ for Cases A--E\@.  The solid black contours represent the magnetic flux $A_z$.\label{bzniplot}}
\end{figure}

\begin{figure}[tp]
    \begin{center}
    \includegraphics[width=3.3in]{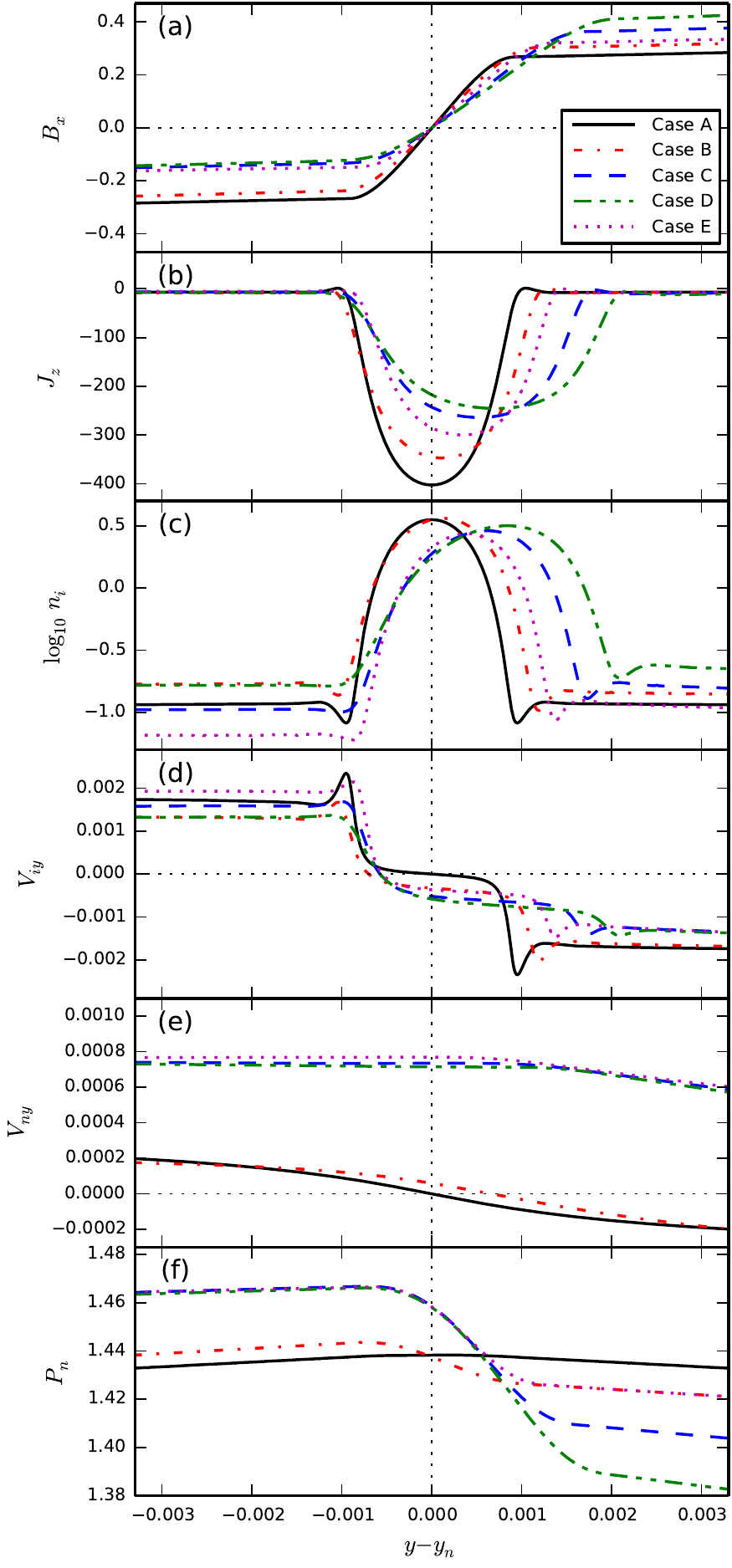}
    \end{center}
    \caption{
        The structure of the current sheet along $x=0$ at $t=25$ very near the current sheet relative to the X-point.  Shown are (a) the reconnecting component of the magnetic field $B_x$, (b) the out-of-plane current density $J_z$, (c) the logarithm of ion density $\log_{10}\,n_i$, (d) the inflow component of ion velocity $V_{iy}$, (e) the inflow component of neutral velocity $V_{ny}$, and (f) the neutral pressure.
    \label{inflowslice}}   
\end{figure}

All five simulations show broadly similar evolution.  The electric field application from $t=0$ to $t=5$ described in Section \ref{initialization} allows the inflow/outflow pattern associated with two-dimensional reconnection to develop in the portion of the current sheet near the origin.  The outflow jet lengthens as it plows into a region of enhanced ion density associated with current sheet thinning outside of the reconnection region.  By $t\sim 25$, laminar reconnection is well-established.  The current sheet has thinned from its initial thickness of ${0.1}$ to a thickness of $\delta\sim 1.0\times 10^{-3}$ ({$\sim$}$10$ m).  This is comparable to the value of the neutral-ion mean free path evaluated inside the current sheet: $\lambda_{ni} \gtrsim 6.0\times 10^{-4}$ to $1.0\times 10^{-3}$ ({$\gtrsim$}$6$ to $10$ m).  The ionization fraction inside each current sheet is of order $0.01$ at this time.  The structure of the reconnection region for all simulations at $t=25$ is shown in Figures \ref{jzvxplot}, \ref{flowplot}, \ref{bzniplot}, and \ref{inflowslice}.  At $t=25$, the current sheets have thinned to {$\sim$}$\lambda_{ni}$, laminar reconnection is well-established, and plasmoid formation has not yet begun.  The plasmoid instability onsets during all of these simulations.  The simulations end when structures develop on scales comparable to the resolution scale as a result of the plasmoid instability.

Figure \ref{jzvxplot} shows the out-of-plane current density and the ion outflow.  In Case B, the current sheet is slightly arched so that the X-point is closer to the low temperature upstream region ($y>0$) than the ends of the current sheet.  In the cases with magnetic asymmetry, the current sheet is arched so that the X-point is closer to the strong field upstream region than the ends of the outflow jets (see also Fig. \ref{inflowslice}b).  Case E, which has the strong field side coincident with the high temperature side, shows the fastest development of these three cases and the strongest current density.

As in \citetalias{leake:partial1} and \citetalias{leake:partial2}, the ion and neutral outflows are tightly coupled.  The right half of each panel in Figure \ref{jzvxplot} only show pseudocolor maps of $V_{ix}$ because $V_{nx}$ is very similar in the outflow jet.  The strong coupling of the ion and neutral outflow jets occurs because the mean free path for neutral-ion collisions is comparable to the current sheet thickness which is significantly shorter than the length of the current sheet.  

Figure \ref{flowplot} shows the inflow components of the ion velocity on the left side of each panel and the neutral inflow speed on the right side (see also Figs.\ \ref{inflowslice}d and \ref{inflowslice}e).  As in \citetalias{leake:partial1} and \citetalias{leake:partial2}, the inflow velocities are decoupled. In Cases B--E, the decoupling is asymmetric in part because $\lambda_{ni}$ is different in each upstream region.  The neutrals and the ions at the X-point are moving in opposite directions.  For simulations with magnetic asymmetry (Cases C, D, and E), there is neutral flow through the current sheet from the weak magnetic field side toward the strong magnetic field side.  This corresponds to a neutral pressure gradient that is pushing neutrals from the weak field side into the strong field side (see Fig.\ \ref{inflowslice}f).  

The ion density is strongly peaked within the current sheet for all cases, as shown on the left side of each panel in Fig.\ \ref{bzniplot} \citepalias[see also][]{leake:partial1, leake:partial2}.  The decoupling of ions and neutrals on scales below the neutral-ion mean free path allows the ions to be swept into the current sheet by the magnetic field.  Neutrals are dragged along by collisions with ions, though less effectively on these short length scales.  The current sheet is therefore out of ionization equilibrium: the ionization fraction is much higher than the equilibrium value.  As a result, recombination becomes of comparable importance to the outflow in the ion continuity equation.

Next we consider the consequences of changing the temperature asymmetry while maintaining the same magnetic asymmetry.  As we go from D to C to E, the magnetic asymmetry remains constant ($\basym=0.5$) but the temperature asymmetry goes from $\tasym=1.06$ to $1$ to $0.95$.  Case D has higher temperature in the weak field upstream region; Case C has initially uniform temperature; and Case E has higher temperature in the strong field upstream region.  

\begin{figure}[bt]
    \begin{center}
        \includegraphics{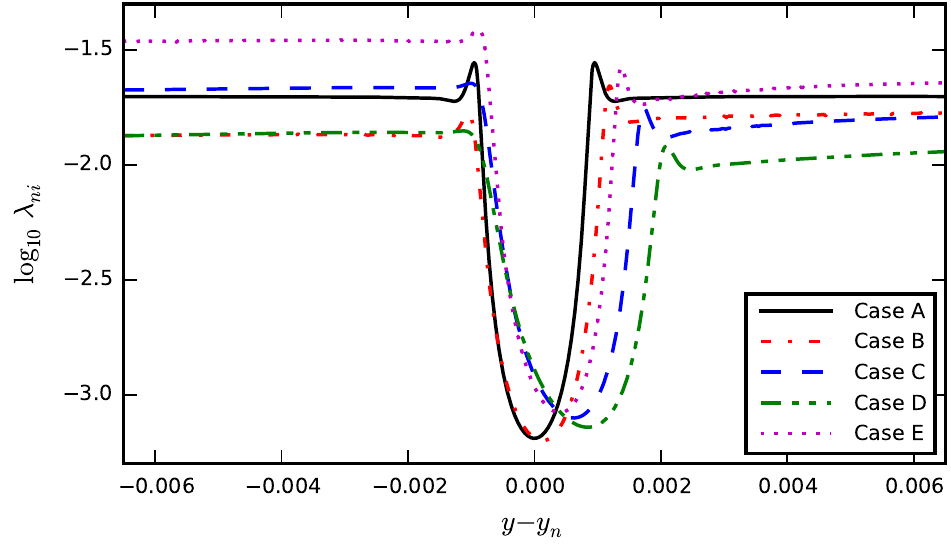}
    \end{center}
    \caption{The logarithm of the neutral-ion mean free path $\lambda_{ni}=V_{T,n}/\nu_{ni}^\dagger$ along $x=0$ at $t=25$. The minimum values of $\lambda_{ni}$ in this plot are $6.5\times 10^{-4}$, $6.3\times 10^{-4}$, $7.9\times 10^{-4}$, $7.2\times 10^{-4}$, and $8.3\times 10^{-4}$ for cases A through E, respectively.
        \label{lambda_ni}}
\end{figure}

Switching the temperature asymmetry while maintaining the same magnetic asymmetry has a significant impact on how quickly reconnection develops, the reconnection rate at a given time, and the onset time and mode structure of the plasmoid instability.  Reconnection develops more quickly and occurs at a faster rate in Case E where the strong field upstream region has a higher temperature than the weak field upstream region. As we proceed from Case D to C to E, the relative velocity between the ions and neutrals increases in magnitude on the weak field side but does not change substantially on the strong field side (see, e.g., Fig.\ \ref{inflowslice}d while noting that the $V_{ny}$ profiles in Fig.\ \ref{inflowslice}e are nearly identical between these three runs).  The neutral-ion mean free path $\lambda_{ni}$ along $x=0$ at $t=25$ is shown in Fig.\ \ref{lambda_ni}.  In this figure, the minimum value of $\lambda_{ni}$ increases by 10\% from Case D to C and another 5\% from Case C to E\@.  In contrast, the values of $\lambda_{ni}$ increase by {$\sim$}$30$\% to {$\sim$}$60$\% from D to C to E in both near-upstream regions.  The reconnection rate at $t=25$ measured as $\dif A_z/\dif t$ evaluated at the X-point along $x=0$ is $3.4\times 10^{-4}$, $3.8\times 10^{-4}$, and $4.5\times 10^{-4}$ in Cases D, C, and E, respectively, which suggests that differences in ion-neutral coupling inside and around the current sheet are largely responsible for the observed differences between simulations.  In particular, the reconnection rate in these simulations is greater when there is weaker coupling between neutrals and ions.  The $V_{ny}$ profile in Fig.\ \ref{inflowslice}d remains nearly unchanged between simulations with the same magnetic asymmetry, which suggests that the neutral flow across the current sheet depends strongly on magnetic asymmetry but only weakly on temperature asymmetry.

\subsection{Role of the Hall Effect\label{halleffect}}

The Hall effect is not expected to be significant in the solar chromosphere during magnetic reconnection unless structures develop on scales comparable to or below the ion inertial length $d_i$ \citep{malyshkin:2011}.  In the case of strong coupling between ions and neutrals, it has been proposed that the ion inertial length may be enhanced because each ion will be dragging along a much larger mass \citep{pandey:2008a, malyshkin:2011}; consequently the effective ion inertial length is proposed to be $d_i' = d_i \sqrt{\rho/\rho_i}$.  We therefore include the Hall effect in these simulations. 

The characteristic thickness of the current sheet is {$\sim$}$10^{-3}$ ($10$ m; see Fig.\ \ref{inflowslice}). The ion inertial length in the upstream regions is initially of order $d_i\sim 4\times 10^{-4}$ ($4$ m; using upstream parameters from Case A), while $d_i'\sim 0.02$ ($200$ m).  In the current sheet, the pileup of ions leads to a local ion inertial length of $d_i\approx 8\times 10^{-5}$ ($0.8$ m; based on an ion density of {$\sim$}$3n_0$), which corresponds to $d_i' \sim 8\times 10^{-4}$ ($8$ m).  These lengths scales can also be compared to $\lambda_{ni}$ which is of order {$\sim$}$0.01$ to $0.03$ in the regions just upstream of the current sheet and {$\sim$}$10^{-3}$ in the current sheet.

The right side of Figure \ref{bzniplot} shows the quadrupole structure of the out-of-plane magnetic field $B_z$ that forms as a result of the Hall effect during laminar reconnection.  The quadrupole field strength remains about an order of magnitude smaller than the asymptotic values of $B_x$, but $|B_z|/\sqrt{B_x^2+B_y^2}$ locally reaches {$\sim$}$0.5$ near the location where $|B_z|$ is largest on the weak field upstream region in Cases C--E\@.  This occurs in part because the in-plane field is weaker in the near upstream regions than in the far upstream regions.  Despite the locally high value of the ratio of $|B_z|$ to the in-plane field, these cases show neither the significant enhancement of the reconnection rate nor the development of the low aspect ratio (X-point) geometry expected during Hall-mediated reconnection in fully ionized plasmas \citep[e.g.,][]{biskamp:1997}.

The structure of the quadrupole field is modified by both temperature and magnetic asymmetries.  When the temperature or magnetic asymmetries are changed, this also impacts the neutral density, ion density, and ionization fraction asymmetries and therefore $\lambda_{ni}$.  In Case B, the quadrupole lobe on the low ion density side ($y<0$) has a greater spatial extent and higher magnitude than the high ion density side.  In Cases C--E with magnetic asymmetry, the quadrupole lobe on the strong magnetic field side is localized near the current sheet while the quadrupole lobe on the weak magnetic field side has a greater spatial extent.  This asymmetry of the quadrupole field is qualitatively similar with previous fully kinetic simulations \citep[see, e.g., Fig.\ 8c of][]{pritchett:2008:asym}.  As we proceed from Case D to C to E at $t=25$, the quadrupole field on the strong field side increases in strength while the quadrupole field on the weak field side decreases in magnitude but increases in spatial extent.  

Even though the ion inertial scale is shorter than the current sheet thickness throughout these simulations, this might not be the case during the long-term evolution of the plasmoid instability.  While simulations of fully ionized plasmas have shown that the plasmoid instability allows reconnection to occur at a rate that is roughly independent of Lundquist number, \citet{Daughton:2009} and \citet{shepherd:2010} have proposed that the most important role of the plasmoid instability might actually be to allow structure to develop on scales smaller than the ion inertial length which would then allow fast collisionless reconnection to take place.  This proposed mechanism has not yet been tested in weakly ionized plasmas, which would require capturing highly nonlinear behavior at later times and will be explored in future work.  

\subsection{Plasmoid Formation\label{plasmoidformation}}

\begin{figure}[tp]
    \vspace{-7mm}
    \begin{center}
        \includegraphics[height=7.5in]{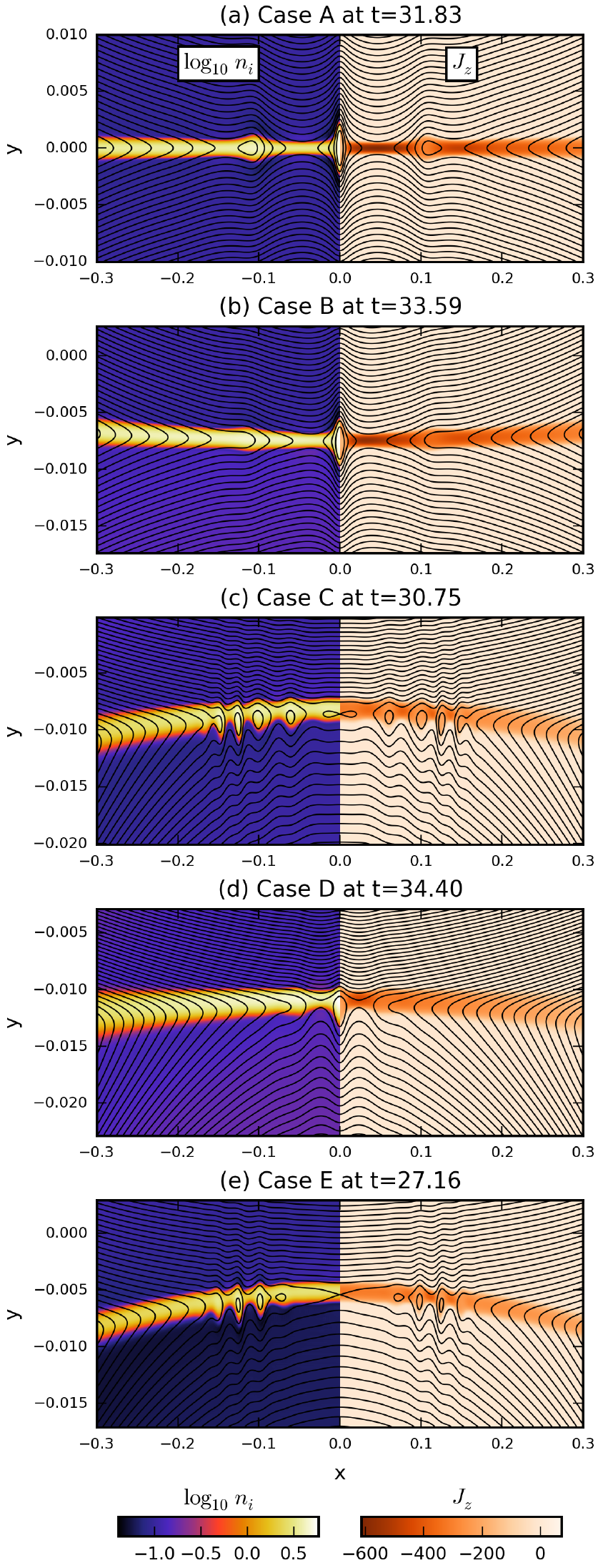}
    \end{center}
    \vspace{-7mm}
    \caption{The ion density $n_i$ (left) and current density $J_z$ (right) for Cases A--E near the end of each simulation after plasmoids form.
        \label{plasmoidplot}}
\end{figure}

Each simulation shows the formation of plasmoids after the current sheets have thinned sufficiently.  Figure \ref{plasmoidplot} shows the ion density, out-of-plane current density, and magnetic flux during the early nonlinear evolution of the plasmoid instability for each case.  In contrast to Figures \ref{jzvxplot}--\ref{lambda_ni}, the times for each panel were chosen to be near the end of each simulation but before the plasma substructures associated with the cascading nonlinear evolution of the plasmoids reached the grid scale. The appearance of additional in-plane null points occurs at $t=27.9$ for Case A, $t=29.8$ for Case B, $t=29.6$ for Case C, $t=31.1$ for Case D, and $t=25.9$ for Case E\@.  In all cases, the secondary islands develop in the central portion of the current sheet: within $x\lesssim 0.16$ compared to a current sheet length of {$\sim$}$0.8$.  The out-of-plane current density has a local extremum near each X-point.

We first consider the effect of introducing a temperature asymmetry by comparing Cases A and B\@.  The plasmoid instability takes longer to develop in Case B which has temperature asymmetry, but the mode structure does not change substantially. Both cases show a central O-point with two X-points on each side with an additional X-point/O-point pair further out.  The symmetry about $x=0$ allows a pitchfork bifurcation to change the central X-point into an O-point.  The resulting island then grows due to reconnection and is unable to be advected out of the reconnection region by the outflow.  Prior simulations of the plasmoid instability in resistive MHD have avoided this situation by simulating the entire domain and including a slight asymmetry along the outflow direction so that any central islands that form do not remain in the current sheet indefinitely.

The structure of the current sheet and the resultant plasmoid instability are more directly modified by magnetic asymmetry, including when it is coupled with temperature asymmetry.  While Case D also undergoes a pitchfork bifurcation to change the central X-point into an O-point, Cases C and E maintain a central X-point and develop multiple alternating X-points and magnetic islands.  As in prior simulations using the resistive MHD approximation \citep{murphy:plasmoidasym}, the resulting islands preferentially grow into the weak field upstream region.  

We now compare our simulations with the complementary simulations of the plasmoid instability during partially ionized chromospheric reconnection by \citet{ni:2015}.  Ni et al.\ capture the long term nonlinear evolution of the plasmoid instability using a single fluid approach that incorporates ambipolar diffusion and assumes ionization equilibrium.  The parameter regimes for the two sets of simulations differ.  For example, Ni et al.\ assume an initial upstream $\beta$ of $0.1$ which allows heating up to {$\sim$}$8\times 10^4$ K, while our simulations have high $\beta$ and thus do not result in significant heating because less magnetic free energy is available.  Ni et al.\ present current sheets with lengths of order {$\sim$}$1$ Mm which is a significant fraction of the size of the chromosphere, in contrast to lengths of {$\sim$}$10$ km in our simulations.  The thicknesses of the Ni et al.\ current sheets are generally much larger than $\lambda_{ni}$ which suggests that the neutrals and ions are strongly coupled in their regime.  While recombination plays an important role in the simulations presented by \citetalias{leake:partial1}, \citetalias{leake:partial2}, and this work, the high temperatures found by Ni et al.\ make it less likely that net recombination will play an important role in the plasma continuity equation.  Physical parameters such as magnetic field strength, temperature, density, and ionization fraction vary significantly within the chromosphere even at a single height, so it is likely that there are regions where each parameter regime is valid.  These differences highlight the need for detailed parameter studies on how partial ionization impacts the reconnection process including the plasmoid instability at various locations in the chromosphere.

\subsection{Reconnection Rate\label{reconrate}}

\begin{figure}[tb]
    \begin{center}
    \includegraphics[width=8.5cm]{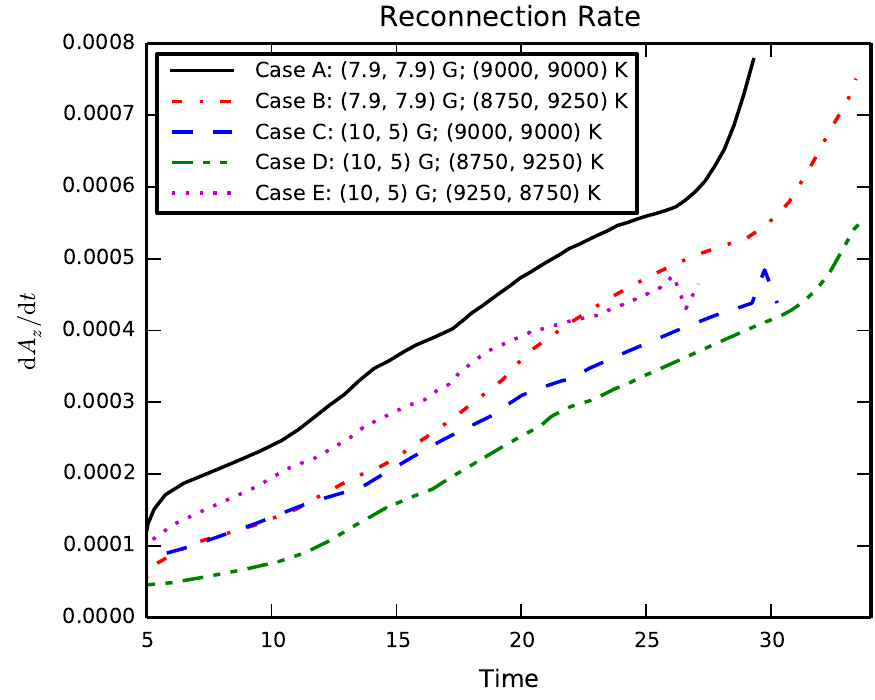}
    \end{center}
    \caption{
        The reconnection rate as a function of time as measured by the maximum change in magnetic flux among all X-points in the current sheet, $\dif A_z/\dif t$.  This rate is presented in dimensionless units according to Section \ref{normalizations}; consequently, this rate has not been renormalized to values immediately upstream of the reconnection layer.
    \label{ratefig}}
\end{figure}

The reconnection rate for all five simulations is shown in Fig.\ \ref{ratefig}.  This rate was measured as the maximum of the time derivatives of $A_z$ at all X-points within the reconnection region.  A common alternative method for measuring the reconnection rate is to measure the inflow velocity divided by the outflow velocity; however, during asymmetric reconnection there is ambiguity in how to define the inflow velocity.  The reconnection rate before $t=5$ is not shown because an electric field was applied to initialize reconnection at early times.  When interpreting this figure, it is important to recall that the initial total pressure and average upstream magnetic energy density are constant between runs.  With this convention, magnetic and temperature asymmetry reduce the reconnection rate for all cases studied.

From early times until around when plasmoids form, the reconnection rate is highest at the X-point along $x=0$.  The reconnection rate increases as the current sheet thins.  Around the time that plasmoids form, the reconnection rate at this X-point decreases.  For some cases, this X-point bifurcates into an O-point.  The peak reconnection rate then occurs at one of the nearby X-points that is not located along $x=0$.  Cases A, B, and D show a considerable increase in the reconnection rate after the formation of plasmoids.  It is likely that Cases C and E would also show a similar increase in the reconnection rate; however, the simulations ended before this occurred.  At even later times, structure on small enough scales could develop so that Hall reconnection could become important \citep[see also][]{Daughton:2009, shepherd:2010}.

\subsection{X-Point Motion and Flows Across the X-Point\label{nullflows}}

\begin{figure*}[t!]
    \begin{center}
    \includegraphics[width=6.5in]{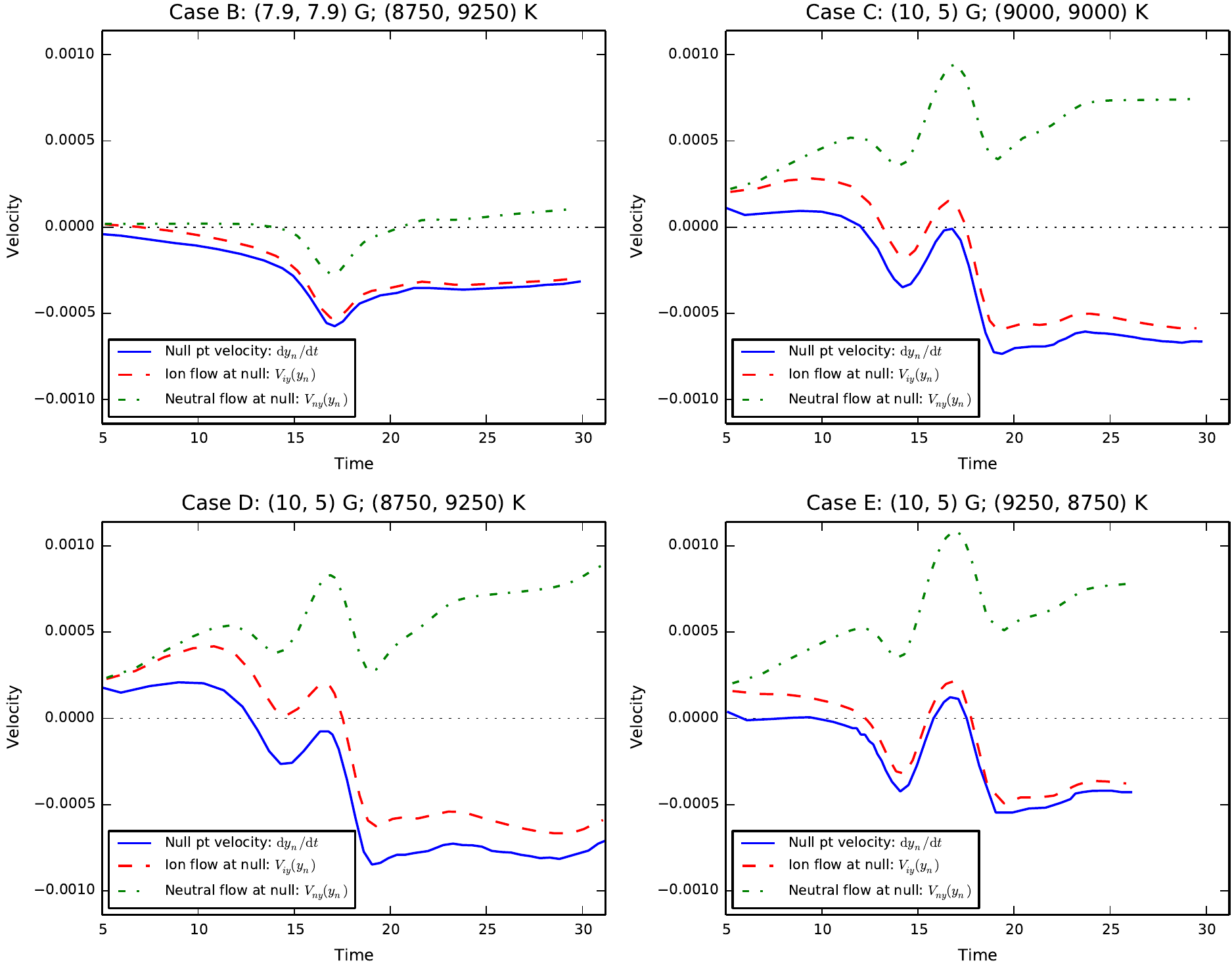}
    \end{center}
    \caption{A comparison between the X-point velocity $\frac{\dif y_n}{\dif t}$, the ion flow at the X-point along the inflow direction $V_{iy}(y_n)$, and the neutral flow at the X-point $V_{ny}(y_n)$ for Cases B--E\@.  We consider only the X-point along $x=0$ and times after the early electric field application and before the onset of the plasmoid instability.  Case A is not shown because all of these velocities equal zero due to symmetry.
    \label{nullflowfig}}  
\end{figure*}

We consider three velocities that are relevant to the small-scale physics near the X-point located along the symmetry axis at $\mathbf{x}_n = (0,y_n)$.  The first velocity is the rate of motion of the X-point along the inflow direction: $\frac{\dif y_n}{\dif t}$.  This velocity is not a fluid velocity, but rather the velocity of a topologically stable magnetic feature \citep{murphy:retreat}.  The second is the ion flow at the X-point along the inflow direction: $V_{iy}(y_n)$.  The third is the neutral flow at the X-point along the inflow direction: $V_{ny}(y_n)$.  Differences between $\frac{\dif y_n}{\dif t}$ and $V_{iy}(y_n)$ must be due to non-ideal behavior such as resistivity or the Hall effect.  Differences between $V_{iy}(y_n)$ and $V_{ny}(y_n)$ correspond to momentum transfer between ions and neutrals (e.g., Eq.\ \ref{rinidef}).

Figure \ref{nullflowfig} shows $\frac{\dif y_n}{\dif t}$, $V_{iy}(y_n)$, and $V_{ny}(y_n)$ as a function of time for all of the cases with asymmetric inflow.  Case A is not shown because these velocities equal zero due to symmetry.  The bump in velocity around $t=16$ is a consequence of the initial conditions being slightly out of equilibrium; the resulting pulse is mostly but not entirely damped by a viscous layer at the boundary and the reflected pulse returns to the current sheet region at this time.  The magnitude of this pulse is weak compared to the ion inflow speeds, but is the same order of magnitude as the velocities associated with each null point.  For the rest of this section, we consider $t\gtrsim 20$ for which reconnection is well established.  

In general, there is a small but nonzero difference between $\frac{\dif y_n}{\dif t}$ and $V_{iy}(y_n)$.  This similarity indicates that the magnetic field is predominantly carried by the ion flow.  The residual difference in flow results from resistive diffusion of the magnetic field and the Hall effect.  The difference between either of these velocities and the neutral flow at the X-point $V_{ny}(y_n)$ is greater, which is consistent with decoupling between the ion and neutral inflows.  The $V_{ny}$ profile along the inflow direction remains roughly constant between Cases A and B which both have symmetric magnetic fields but different temperature profiles, and also between Cases C--E which have the same asymmetric magnetic field configuration but differing initial temperature profiles. Case B has symmetric upstream magnetic fields but the $y<0$ upstream region is warmer and less dense than the $y>0$ upstream region.  That $\frac{\dif y_n}{\dif t} \approx V_{iy}(y_n) < 0$ indicates that the X-point is being carried primarily by the ions in the direction of the warmer upstream region.  

The simulations with magnetic asymmetry (Cases C, D, and E) show neutral flow from the weak field upstream region into the strong field upstream region.  This flow results from a neutral pressure gradient that is pushing the neutrals from the weak field side into the strong field side where the neutral pressure is lower.  This flow is able to occur because of imperfect coupling between species on scales below or comparable to the neutral-ion mean free path.  Neutrals swept along with the outflow will have preferentially originated from the weak field upstream region.  This flow may have important observational consequences because the abundances of neutrals swept along with the outflow are likely to better match the weak field upstream region.  In contrast, there is net ion flow inside the current sheet from the strong field upstream region into the weak field upstream region.  While the neutrals are pushed by a neutral pressure gradient, the dominant force on the ions is a magnetic pressure gradient.  

\section{DISCUSSION\label{discussion}}

In this paper we perform simulations of asymmetric magnetic reconnection in partially ionized chromospheric plasmas using the plasma-neutral module of the HiFi framework.  We include cases with symmetric or asymmetric upstream temperatures and magnetic field strengths.  The plasma and neutrals are modeled as separate fluids with momentum and energy transfer between species due to collisions and charge exchange.  These simulations self-consistently include ionization and recombination without assuming ionization equilibrium.  

Several of the properties of asymmetric partially ionized reconnection are qualitatively similar to the symmetric cases \citepalias[Case A;][]{leake:partial1, leake:partial2}.  The current sheet rapidly thins so that its thickness becomes comparable to the neutral-ion mean free path evaluated inside the current sheet.  The ion and neutral outflows are strongly coupled, but the ion and neutral inflows are decoupled.  The reconnecting magnetic field drags ions into the current sheet which leads to a significant enhancement of ion density so that the current sheet is out of ionization equilibrium.  Recombination becomes of comparable importance to the outflow in the equation for ion conservation of mass.  In this paper, we show that these properties are modified by the temperature and magnetic field asymmetries.

During our simulations of asymmetric reconnection, the ion and neutral inflows remain decoupled, but the decoupling is asymmetric.  When there is magnetic asymmetry, there is net neutral flow through the current sheet from the weak magnetic field upstream region into the strong field upstream region.  This neutral flow results from a large scale neutral pressure gradient and imperfect coupling between ions and neutrals along the inflow direction.  Similarly, the greater Lorentz force acting on the ions from the strong field upstream region leads to the ions pulling the X-point into the weak field upstream region.  These effects are not present during symmetric simulations.  An observational consequence of these neutral flows through the current sheet is that most of the neutrals swept along with the outflow are more likely to have originated from the weak magnetic field upstream region.  This may be especially important if there are different elemental abundances in each upstream region (e.g., different FIP enhancements).  

Prior simulations of asymmetric reconnection in fully ionized plasmas have shown non-ideal flows through null points \citep[e.g.,][]{oka:2008, murphy:retreat, murphy:double}.  These flows result from non-ideal effects such as resistivity and the Hall effect.  In resistive MHD, for example, the motion of null points results from a combination of resistive diffusion of the magnetic field and advection by the bulk plasma flow \citep{murphy:retreat}.  During asymmetric reconnection in partially ionized plasmas, an additional contributor to non-ideal plasma flows across null points is the electric field contribution from electron-neutral drag (see the fourth term on the right hand side of Eq.\ \ref{genohms}).  Neutral flow through the current sheet is a new effect that is not present in fully ionized situations.

We include the Hall effect in our simulations to test whether or not a transition to Hall reconnection can develop in this regime.  We find that the out-of-plane quadrupole magnetic field associated with antiparallel Hall reconnection does develop.  The out-of-plane magnetic field strength is a significant fraction of the in-plane field strengths just upstream of the current sheet.  However, the characteristic low aspect ratio geometry and high reconnection rate expected during Hall reconnection do not develop over the course of these simulations, including during the early nonlinear evolution of the plasmoid instability.  The transition to Hall reconnection in a partially ionized plasma has been investigated analytically by \citet{malyshkin:2011}, and will continue to be investigated by ongoing simulation efforts to determine the conditions under which such a transition is likely to occur.

After an initial phase, each of our simulations show the development of the plasmoid instability.  These simulations capture the early nonlinear evolution until structures develop on scales comparable to the resolution scale.  In cases with magnetic asymmetry, the plasmoids develop preferentially into the weak field upstream region, which is similar to fully ionized simulations. Recombination due to the enhanced ion density remains an important loss term in the plasma continuity equation.  We anticipate that secondary merging of plasmoids will be further modified by these asymmetries.  Our simulations complement the work of \citet{ni:2015} who use a single fluid framework to investigate the long term nonlinear evolution of the plasmoid instability in a different parameter regime.

We investigate the coupling between magnetic and temperature asymmetries by comparing cases with the same magnetic asymmetry but different temperature asymmetries.  We find that reconnection develops more quickly, occurs at a faster rate, and becomes unstable to plasmoid formation as we increase the temperature in the strong field upstream region while decreasing the temperature in the weak field upstream region.  This change corresponds to $\lambda_{ni}$ increasing in both upstream regions.  The neutral flow through the current sheet does not noticeably change and therefore depends more strongly on the magnetic asymmetry than the temperature asymmetry.

There are many important remaining directions for modeling work to understand how reconnection occurs in the solar chromosphere and other weakly ionized plasmas.  The simulations of \citetalias{leake:partial1}, \citetalias{leake:partial2}, and this paper have all simulated two-dimensional, antiparallel reconnection.  The simulations presented in these works have modeled the early nonlinear evolution of the plasmoid instability in a weakly ionized plasma, and to model the later evolution will likely require a combination of increased resolution (including along the outflow direction to capture plasmoid merging) and increased diffusion.  An alternative strategy would be to evolve the logarithm of density instead of the density itself \citep[e.g.,][]{elee:2014} which precludes the possibility of the number density becoming negative.  \citet{ni:2015} adopt adaptive mesh refinement to ensure adequate resolution.  The presence of a guide field can suppress the thinning of current sheets due to ambipolar diffusion \citep{brandenburg:1994, brandenburg:1995}, and thus should be considered in future efforts \citep[see also][]{ni:2015}.  Physical conditions in the chromosphere vary significantly even at a single height, so a detailed examination of parameter space will be necessary to characterize the different regimes of partially ionized reconnection in the chromosphere.  The parameter studies may also include regimes relevant to partially ionized plasmas in the laboratory and in astrophysics.  Future studies should investigate the impact of a realistic geometry and the interplay between small-scale physics and global dynamics (e.g., how small and large scales feed back on each other).  Finally, much remains to be understood about the role of three-dimensional effects during weakly ionized reconnection and the behavior of current sheet thinning in cases with and without null points. 

In addition to the modeling efforts, much work remains to compare the simulation results to observations of chromospheric reconnection and to validate the results against laboratory experiments on partially ionized reconnection.  A challenge in comparing simulations against observations is the disparity in length scales.  The current sheets in these simulations have characteristic thicknesses of $\lesssim 10^2$~m and lengths of $\lesssim 10$~km, while the spatial resolution of \emph{IRIS} observations is of order $200$~km, so direct diagnostics of the reconnection layer itself remain extremely challenging.  However, the simulations could be used to predict velocities, spectra (including the charge state distributions of minor ions), densities, and morphologies that could then be compared against observation.  The experiments that have recently been performed at MRX provide an opportunity to directly validate HiFi's plasma-neutral module against laboratory measurements where key quantities can be observed using \emph{in situ} probes.  Using a code that has been validated against experimental data will improve confidence that it is able to accurately capture the relevant physical processes in the lower solar atmosphere and in astrophysical plasmas.  

\acknowledgments

The authors thank H.\ Ji, J.\ Leake, J.\ Lin, L.\ Ni, N.\ Nishizuka, J.\ Raymond, K.\ Reeves, C.\ Shen, H.\ Tian, and E.\ Zweibel for useful discussions and an anonymous referee for useful comments that helped to improve this paper.  N.A.M.\ acknowledges support from NASA grants NNX12AB25G, NNX15AF43G, \& NNX11AB61G; NSF SHINE grants AGS-1156076 and AGS-1358342; contract 8100002705 from LMSAL; and NASA contract NNM07AB07C to the Smithsonian Astrophysical Observatory (SAO). V.S.L.\ acknowledges support from the NASA LWS \& Solar and Heliospheric Physics programs, as well as the National Science Foundation. Resources supporting this work were provided by the NASA High-End Computing Program through the NASA Advanced Supercomputing Division at Ames Research Center.  We thank J.\ Sattelberger formerly from SAO and J.\ Chang from NASA for technical support.  This work has benefited from the use of NASA's Astrophysics Data System.

\bibliographystyle{apj}

\end{document}